\documentclass[a4paper,cleveref, autoref, thm-restate,authorcolumns]{article}
\usepackage[dvips]{graphicx}
\usepackage{here}
\usepackage{enumerate}

\title{Multi-species consensus network of DNA strand displacement for concentration-to-strand translation}

\author{
Toshiyuki Yamane\footnote{corresponding author, tyamane@jp.ibm.com}, 
Eiji Nakamura\footnote{current affiliation: enaka@g.ucla.edu, Department of Mechanical and Aerospace Engineering, University of California, Los Angeles, USA}
 and 
Koji Masuda\footnote{masudak@jp.ibm.com}
\\
{\it IBM Research - Tokyo, Kawasaki-shi, Kanagawa-ken 212-0032, Japan}}

\begin{document}

\maketitle

\begin{abstract}
We propose novel chemical reaction networks to translate levels of concentration into unique DNA strand species, which we call concentration translators. 
Our design of the concentration translators is based on combination of two chemical reaction networks, consensus network and conversion network with any number of chemical species.
We give geometric analysis of the proposed CRNs from the viewpoint of nonlinear dynamical systems and show that the CRNs can actually operate as translator.
Our concentration translators exploit DNA strand displacement (DSD) reaction, which is known for a universal reaction that can implement arbitrary chemical reaction networks. 
We demonstrate two specific types of concentration translators (translator A and B) with different switching behavior and biochemical cost and compared their characteristics computationally. 
The proposed concentration translators have an advantage of being able to readout the concentration of targeted nucleic acid strand without any fluorescence-based techniques. 
These characteristics can be tailored according to requirements from applications, including dynamic range, sensitivity and implementation cost.
\end{abstract}

\begin{center}
{\bf keywords}
\end{center}
chemical reaction network, consensus network, DNA strand displacement, concentration translator, heteroclinic orbits, nonlinear dynamical systems

\section{Introduction}
Synthetic biologists have created a variety of artificial biological circuits (hereinafter simply called synthetic circuit), including logic gates, analog circuits, toggle switches, oscillators, and signal amplifiers \cite{Siuti}\cite{Daniel}\cite{Gardner}\cite{Elowitz}\cite{Bonnet}. While some of those circuits have been well-established as tools for fundamental science, therapeutic and diagnostic applications, the list of the biological circuit component still needs to be further extended to realize more complex functions.  

In the present work, we focus on synthetic circuits which act up on the concentration of nucleic acids. 
Computations involving concentrations of target nucleic acids enables us to monitor biological conditions and to diagnose diseases because the concentration of nucleic acids, especially RNA, has meaningful information in living cells.
However, the concentration, which is an analog signal, is not readily accessible because the readout of the concentration usually requires fluorescence-based biochemical techniques which requires expensive tools and time-consuming labors. 
This property is undesirable for extending application areas of synthetic circuits beyond laboratory-scale experiments such as synthetic circuits that can handle the analog signals for mobile and personal healthcare devices.

\begin{figure}[htbp]
\begin{center}
\includegraphics[width=1.0\hsize]{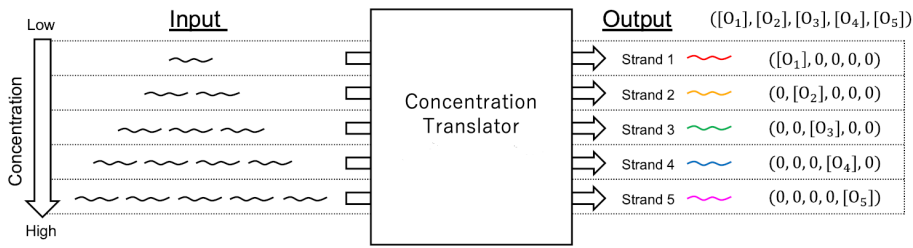}
\caption{Schematic illustration of the function of the concentration translators.}
\label{schematic}
\end{center}
\end{figure}

So far, various analog synthetic circuits have been studied in \cite{Daniel}\cite{QianNature}\cite{Song}, and the characteristics of the analog synthetic circuits investigated theoretically in \cite{Sarpeshkar2014}.
Here, we take a different approach from those of previously reported analog synthetic circuits. 
We propose chemical reaction networks (CRNs) to translates levels of analog-valued concentration into unique DNA strand species. 
We call these CRNs "concentration-to-strand translators" (or simply translators). 
A schematic illustration of the translators is shown in Fig.\ref{schematic}. 
More specifically, we realize the translators as chemical reaction network of higher dimensional consensus networks and conversion networks (or converters).
However, theoretical understanding of consensus networks is insufficient so far since the present consensus networks are limited to small number of species \cite{Chen}.
Therefore, we give geometric analysis from the viewpoint of nonlinear dynamical systems to understand the mechanism of the proposed translators.

Our concentration translators can be implemented biologically by DNA strand displacement (DSD) reactions. 
DSD is a versatile reaction which can implement arbitrary chemical reaction networks with many types of both digital and analog functions \cite{QianNature}\cite{Song}\cite{Soloveichik}\cite{Chen}\cite{QianScience}\cite{Wilhelm}. 
As a proof-of-concept, we give in-silico demonstration of two types of concentration translators, translator A and B in Section \ref{in-silico}. 
Translator A is a composite of two networks; one-way sequential reactions, which we call upconverters, and consensus networks \cite{Chen} and analyze in Section \ref{geometry}. 
Translator B is composed of upconverters and downconverters. 
We numerically analyze dynamics of the two translators and compare their characteristics. 

\section{Geometrical analysis of concentration translator}\label{geometry}
This section describes the concentration translator with arbitrary dimension as combination of concensus network and conversion network, from viewpoint of geometrical theory of nonlinear dynamical systems.
We analyze those two networks separately in subsection \ref{consensus} and subsection \ref{upconversion} and then describe how the combination of these two netwokrs can operate as an translator in subsection \ref{combination}.

\subsection{Consensus network}\label{consensus}
\begin{figure}[htbp]
\begin{center}
\includegraphics[width=0.49\hsize]{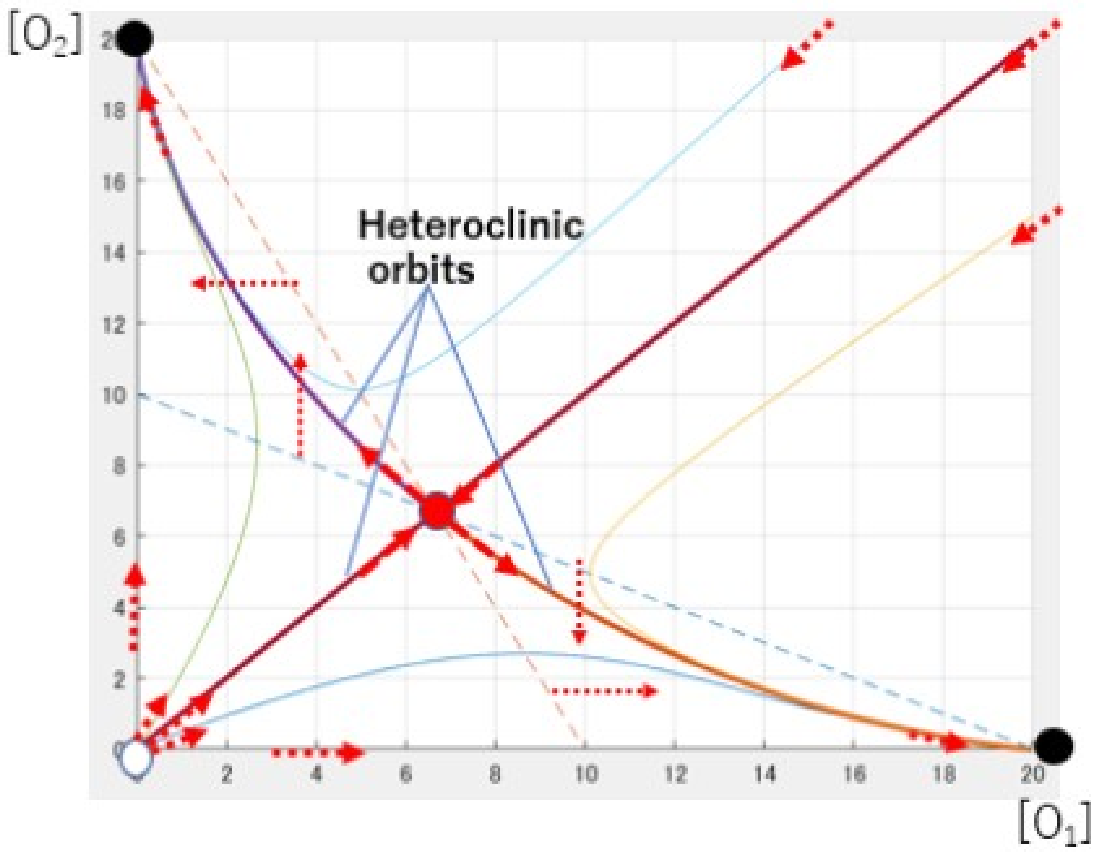}
\includegraphics[width=0.49\hsize]{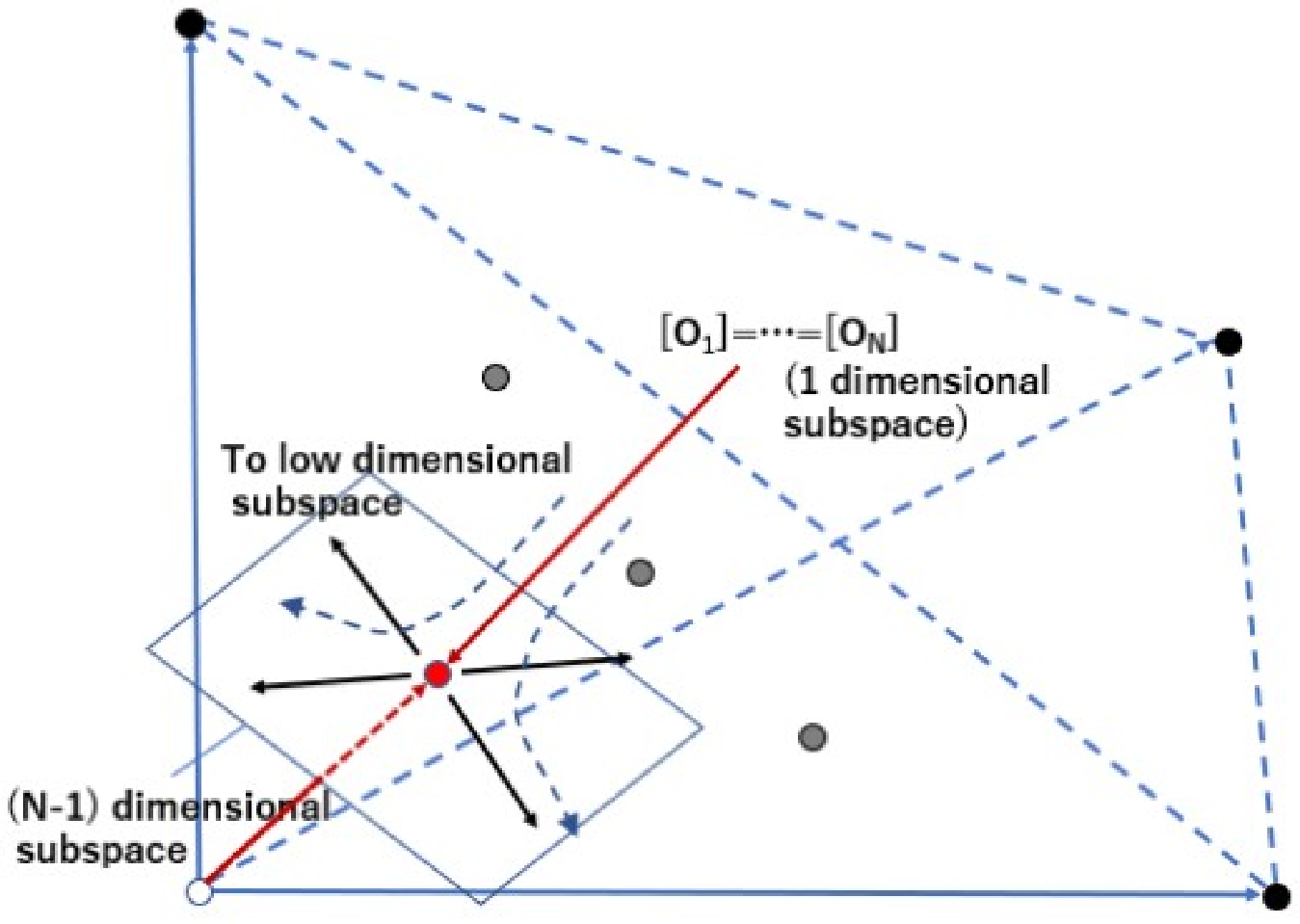}
\caption{(left) Phase portrait of 2-species consensus network. $K=20$. The dashed lines and axes are nullclines. 
The dashed arrows show the vector field on the nullclines.
(right)$2^N$ fixed points (white: the origin, black: stable points on the vertices, gray: saddle points on the faces) and local picture of behaviour around the fixed point $[O_1]=\ldots=[O_N]=K/(2N-1)$ (red).
}
\label{cn}
\end{center}
\end{figure}

The process of consensus formation has been of practical interest in some research areas such as distributed computing and sensor networks. 
For example, consensus on complete graphs was described in \cite{Perron}, where each node has binary (or ternary) states, for example, 1 for yes, 0 for no (and e for undecided). 
After passing the states among the nodes, the network reaches consensus, depending on the initial fraction of the states. 
Later, the consensus network using chemical reaction systems of DNA strand displacement was introduced in \cite{Chen}. 
Their consensus network is formally given by the following chemical reaction system with two main chemical species $O_1$ and $O_2$ as follows:
\begin{eqnarray*}
O_1 + O_2 &\to& 2 X\\
O_1 + X &\to& 2 O_1\\
O_2 + X &\to& 2 O_2,
\end{eqnarray*}
where $X$ denotes a secondary buffer chemical species.
However, their models and analysis has been limited to this two dimensional case and the properties and structures of the system with multi-states remain to be investigated due to the nonlinearity of the system.
For nonlinear systems, one cannot generally hope to find analytical solutions in an explicit way. 
Nonetheless, geometrical qualitative analysis can very often provide us with useful insight on the behavior of the systems \cite{GH}, and  we will perform such kind of analysis for the multi-species consensus networks. 
We start with the following rate equation of 2-species consensus network given by
\begin{eqnarray}
\frac{d\left[O_1\right]}{dt} &=& \left[O_1\right]\left[X\right]-\left[O_1\right]\left[O_2\right],\\
\frac{d\left[O_2\right]}{dt} &=& \left[O_2\right]\left[X\right]-\left[O_1\right]\left[O_2\right],\\
\frac{d\left[X\right]}{dt} &=& 2\left[O_1\right]\left[O_2\right]-\left[O_1\right]\left[X\right]-\left[O_2\right]\left[X\right],
\end{eqnarray}
where $[O_1], [O_2]$ and $[X]$ describe concetrations of corresponding chemical species.
We set the reaction constants to be 1 for simplicity. 
Though the system involves three variables, we can eliminate the secondary variable $\left[X\right]$ and reduce them to two dimensional system using the mass conservation law $\left[O_1\right]+\left[O_2\right]+\left[X\right]=K$ as follows:
\begin{eqnarray}
\frac{d\left[O_1\right]}{dt}&=&\left[O_1\right](K-\left[O_1\right]-2\left[O_2\right]),\\
\frac{d\left[O_2\right]}{dt}&=&\left[O_2\right](K-2\left[O_1\right]-\left[O_2\right]).
\end{eqnarray}
There are four fixed points of the reduced system; 
\begin{equation}
([O_1],[O_2])=(0,0),(0,K),(K,0),(K/3,K/3).
\end{equation}
The eigen value analysis at these four fixed points show that the fixed points $(0,K)$ and $(K,0)$ are stable, and the origin (0,0) are unstable. 
On the other hand, $([O_1],[O_2])= (K/3,K/3)$ is a fixed point of saddle type since the eigen values of the Jacobian at $(K/3,K/3)$ are $-K/3,K/3$ and the corresponding eigen vectors are (1,1) and (-1,1) which are orthogonal to each other.
In addition to the eigen value analysis, nullclines help us understand the dynamics of nonlinear systems. 
A nullcline of a variable is defined as a set of points in the phase space on which the derivative of the variable vanishes. 
When an orbit goes across a nullcline, the sign of the derivative of the variable for the nullcline changes and therefore the nullclines tell us a rough picture of the system behavior. 
In our case, the nullclines are composed of the following four straight lines:  $[O_1]=0$ and $[O_1]+2[O_2]=K$ for $[O_1], [O_2]=0$ and $2[O_1]+[O_2]=K$ for $[O_2]$.
Summarizing all these calculations, we can draw the phase portrait as shown in Fig \ref{cn}(left). 
The line $[O_1]=[O_2]$ separates the phase space into two regions, and we can see the system can operate as consensus network. 
The remarkable feature of the system is that the existence of the orbits connecting two fixed point with two different properties, i.e, saddle and stable/unstable fixed points, which is called heteroclinic orbit. 
The existence of heteroclinic orbits characterizes the overall structure of consensus network because all orbits behave like these heteroclinic orbits.

This geometric analysis can be extended to the following consensus networks with $N$-species. 
\begin{eqnarray}
\frac{d[O_i]}{dt} &=& [O_i]([X]-\sum_{j\neq i}[O_j])\\
\frac{d[X]}{dt} &=& 2\sum_{i,j,i\neq j}[O_i][O_j] - [X]\sum_{i}[O_i].
\end{eqnarray}
Similar to the two-species case, using the law of mass conservation $[O_1]+\ldots+[O_N]+[X]=K$, we have
\begin{equation}
\frac{d[O_i]}{dt} = [O_i](K-[O_i]-2\sum_{j\neq i}[O_j]).
\end{equation}
The phase space of consensus network with $N$ species is a hyper tetrahedron in $N$ dimensional Euclidean space, $[O_i]\ge 0, [O_1]+\ldots+[O_N]\le K$. 
We have two choices of the nullclines $[O_i]=0$ or $K-[O_i]-2\sum_{j\neq i}[O_j]=0$ for each $[O_i]$ and therefore there are $2^N$ fixed points in the $N$-species consensus network.
The dynamics of multi-species consensus network is characterized by the $2^N$ fixed points on the faces and vertices of the hyper tetrahedron and the heteroclinic orbits connecting them.
The fixed point located inner of the hyper tetrahedron is $[O_1]=\ldots=[O_N]=K/(2N-1)$.
The Jacobian at this fixed point is the circulant matrix generated by $N$ dimensional vector $K/(2N-1)\cdot (N-2, -1,\ldots ,-1)$. 
From the general theory of circulant matrices, we can see that it has only one negative eigen value $-K/(2N-1)$ with the eigen vector $(1,\ldots ,1)$ and the other eigen values are all $K(N-1)/(2N-1)>0$.
Fig.\ref{cn}(right) shows $2^N$ fixed points and the local picture of behaviour around the fixed point $[O_1]=\ldots=[O_N]=K/(2N-1)$.
The orbits move along the 1 dimensional stable subspace and then separated by the $N-1$ dimensional unstable subspace depending on the relative magnitude of $[O_i]$'s and lead to the low dimensional subspace along the heteroclinic orbits.

\begin{figure}[h]
\begin{center}
\includegraphics[width=1.0\hsize]{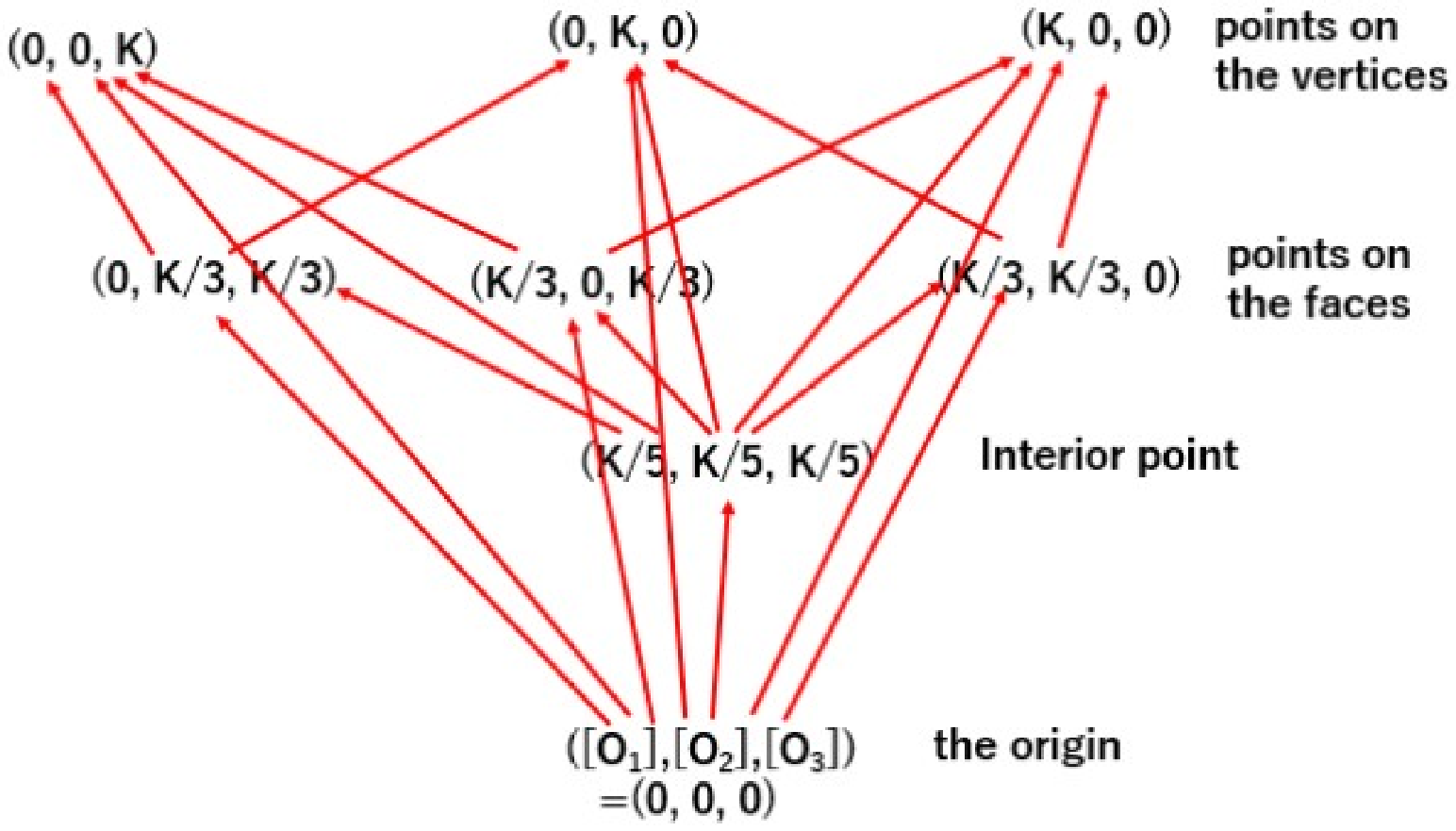}
\includegraphics[width=1.0\hsize]{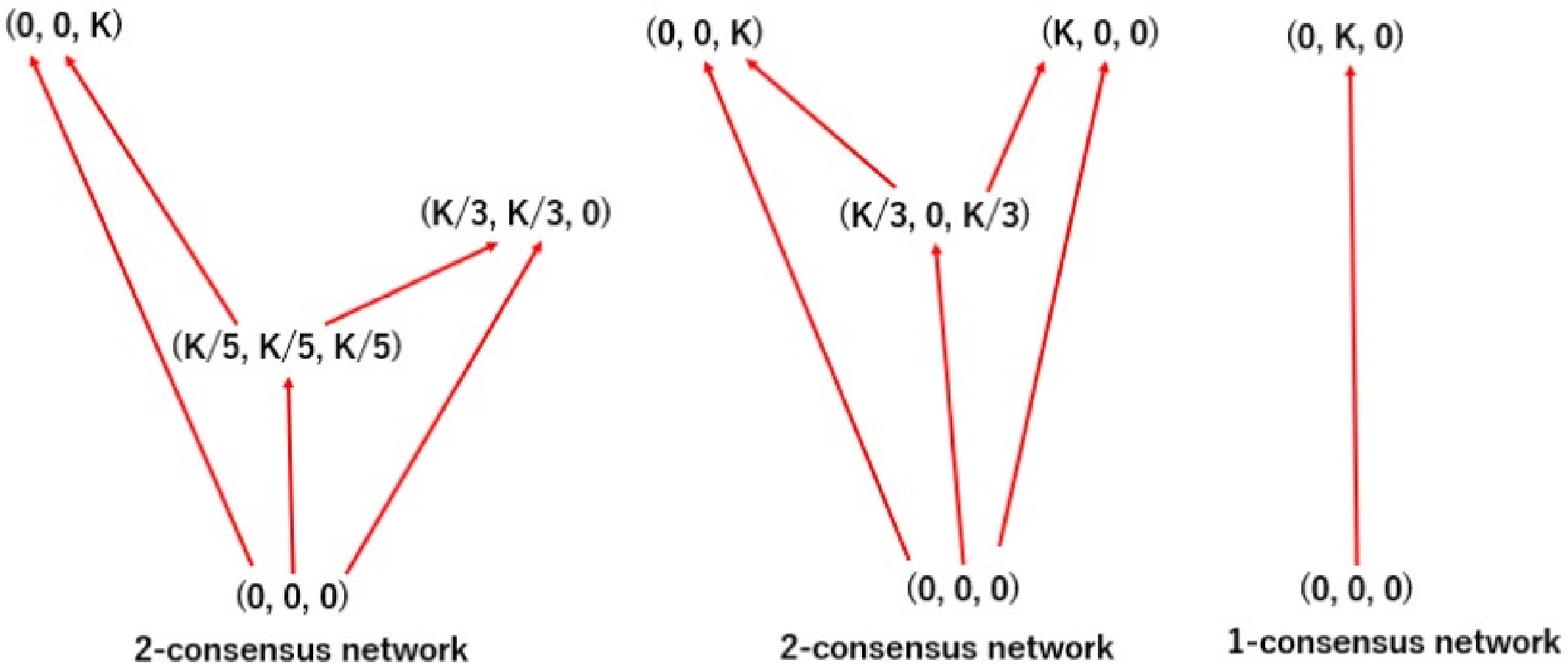}
\caption{(above) Fixed points and heteroclinic orbits of the 3-species consensus network.
(below) sub-consensus networks embedded in a consensus network}
\label{cn3d_hetero_sub}
\end{center}
\end{figure}

Fig.\ref{cn3d_hetero_sub}(above) shows the fixed points and heteroclinic orbits connecting them in the 3-species consensus network. 
Note that a multi-species consensus network naturally contains many sub-consensus networks with fewer species including the trivial consensus network with only one species $[O_i]\rightarrow K$ as shown in Fig.\ref{cn3d_hetero_sub}(below). 
This is because multi-species consensus network reduces to smaller ones if we set some of the variables equal to zero as $[O_i]=[O_j]=\ldots =[O_k]=0$ or set some variables to be equal as $[O_i]=[O_j]=\ldots =[O_k]$. 
The overall dynamics of the consensus network follows one of heteroclinic orbits depending on its initial state and is attracted to lower dimensional subspace. 
Then, the dynamics again follows another heteroclinic orbit of the lower dimensional consensus network embedded in that subspace, and finally reaches one of the stable fixed points on the axis.
In summary, the structure of the multi-species consensus network can be described by hierarchically organized network of heteroclinic orbits.

\subsection{Upconversion network}\label{upconversion}
We define upconversion networks (or simply upconverters) as chemical reaction networks which coonvert one species $O_i$ to next one $O_{i+1}$, in a successive way.
For example, upconversion network with two output species $[O_1]$ and $[O_2]$ is given by
\begin{eqnarray*}
I + G_0 &\to& O_1\\
O_1 + G_1 &\to& O_2,
\end{eqnarray*}
where $[I]$ is a input species, and $G_0$ and $G_1$ are gate species.
The two-species upconversion network is described by the following differential equations.
\begin{eqnarray}
\frac{d [I]}{dt}&=&-[I][G_0],\\ 
\frac{d\left[G_0\right]}{dt}&=&-[I][G_0],\\
\frac{d[G_1]}{dt}&=&-[O_1][G_1], \\
\frac{d\left[O_1\right]}{dt}&=&[I][G_0]-[O_1][G_1],\\
\frac{d\left[O_2\right]}{dt}&=&\left[O_1\right]\left[G_1\right].
\end{eqnarray}
Using the conservation law $[O_1]+[O_2]+[G_0] = G_0(:=[G_0](0)), [O_1]+[O_2]+[I] =I(:=[I](0))$ and $[O_2]+[G_1] = G_1(:=[G_1](0))$, we can eliminate $[I], [G_0]$ and $[G_1]$ and we have
\begin{eqnarray}
\frac{d\left[O_1\right]}{dt} &=& (I-\left[O_1\right]-\left[O_2\right])(G_0-\left[O_1\right]-\left[O_2\right])-\left[O_1\right](G_1-\left[O_2\right]), \\
\frac{d\left[O_2\right]}{dt} &=& \left[O_1\right](G_1-\left[O_2\right]).
\end{eqnarray}

Introducing a new variable $P=\left[O_1\right]+\left[O_2\right]$, we have
\[
\frac{dP}{dt}=(I-P)(G_0-P).
\] 
This is a closed form equation only for $P$, and assuming $G_0>G_1$, we can find easily the final state of $P, [O_1]$ and $[O_2]$ as follows: 
\begin{enumerate}[(a)]
\item If $I<G_1$, then $P\rightarrow I, [O_1] \rightarrow 0, [O_2] \rightarrow I$ and $[I]\to 0$.
\item If $G_1<I<G_0$, then $P\rightarrow I$, $[O_1]\ \rightarrow I-G_1$, $[O_2]\ \rightarrow G_1$ and $[I]\to 0$.
\item If $I>G_0$, then $P\rightarrow G_0$, $[O_1]\ \rightarrow G_0-G_1$, $[O_2]\ \rightarrow G_1$ and $[I]\to I-G_0$.
\end{enumerate}

\begin{figure}[htbp]
\begin{center}
\includegraphics[width=0.8\hsize]{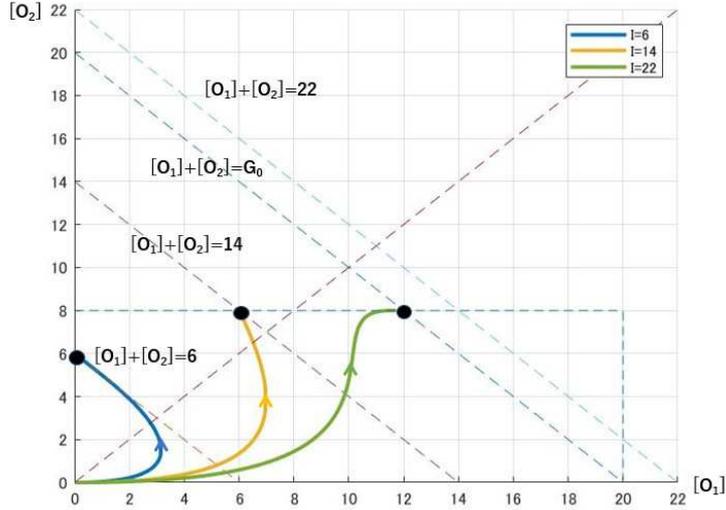}
\caption{Phase portrait of upconversion network. $G_0=20, G_1=8$.}
\label{UN}
\end{center}
\end{figure}

The phase portrait of the upconversion network is shown Fig \ref{UN}. 
In case (a) and (b), the system has a single global fixed point at the intersection of $\left[O_1\right]+\left[O_2\right]=I$ and the edges of the rectangle. 
It moves along the edges while $I$ increases from 0 toward $G_0$. On the other hand, in case (c), it stays at $(G_0-G_1,G_1)$.
The analysis described here can be extended to the higher dimensional upconversion network.
The intersection point moves along the edges of the hyper-cube as $I$ increases from zero, and finally stays at a point on an edge when $I>G_0$.

\subsection{Combining the two networks together}\label{combination}
\begin{figure}[htbp]
\begin{center}
\includegraphics[width=0.4\hsize]{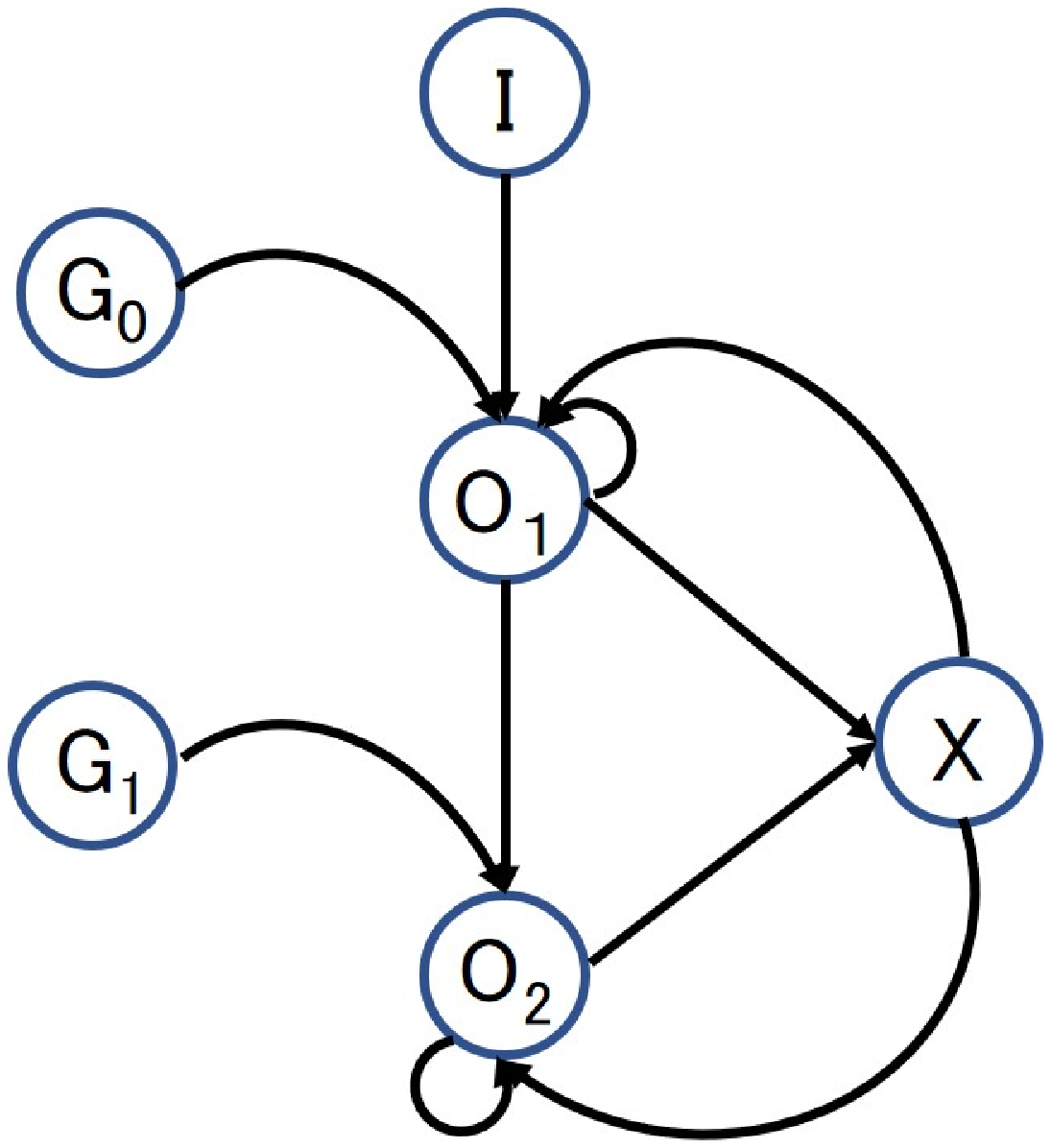}
\includegraphics[width=0.5\hsize]{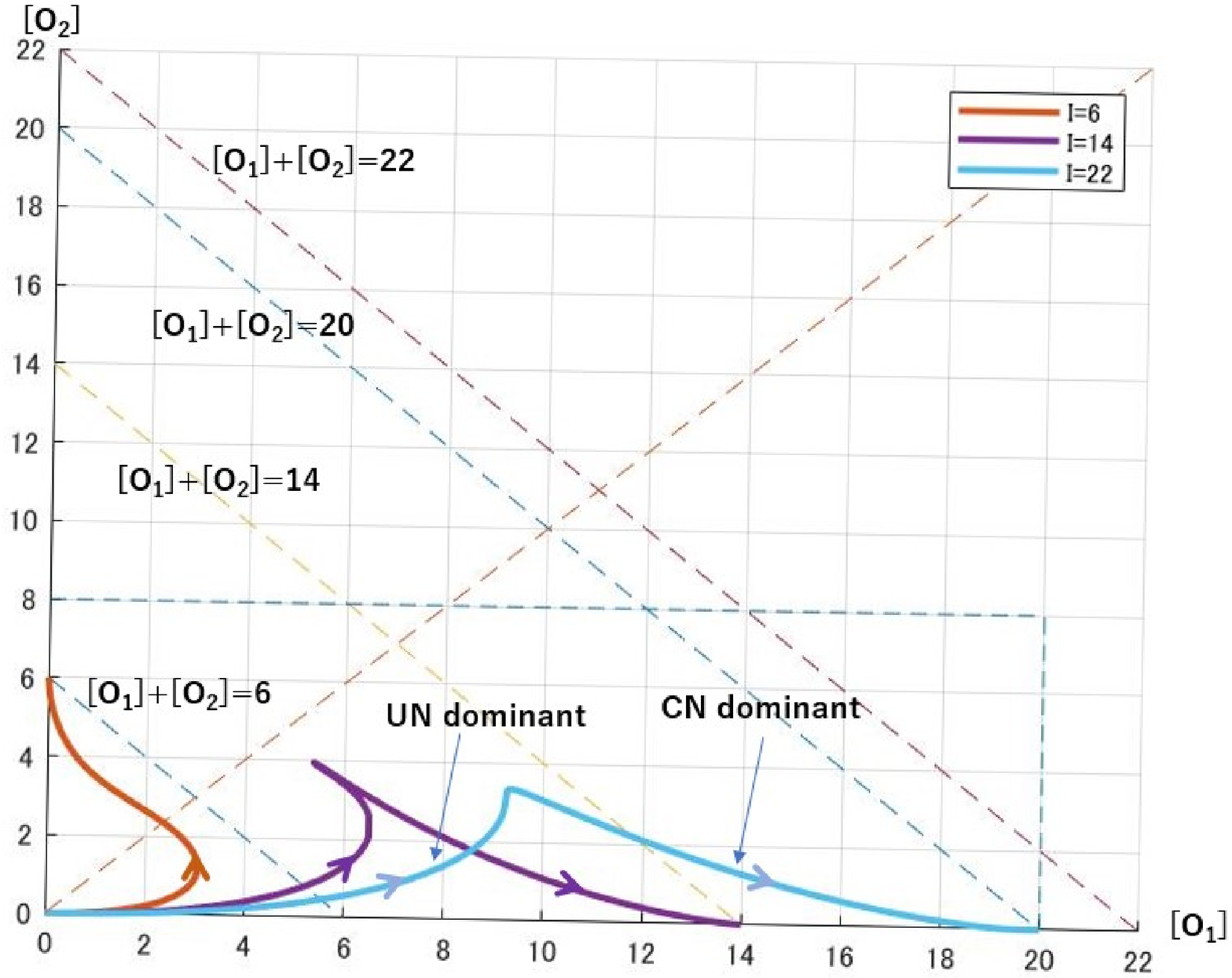}
\caption{(left) Chemical reaction network combining consensus network and upconversion network. (right) phase portrait of consensus network(CN) and upconversion network (UN)}
\label{CN-UN}
\end{center}
\end{figure}

The translator proposed in this paper (see translator A in Section \ref{translator A}) can be understood as collaboration of consensus network and upconversion network.
The simplest chemical reaction network combining two networks can be described in Fig.\ref{CN-UN}(left):
The dynamics is given by the following rate equation involving the six variables $[I],[G_0],[G_1],[O_1],[O_2]$ and $[X]$. 
\begin{eqnarray*}
\frac{d [I]}{dt} &=& -[I][G_0], \\
\frac{d[G_0]}{dt} &=& -[I][G_0]\\
\frac{d[G_1]}{dt} &=& -[O_1][G_1]\\
\frac{d\left[O_1\right]}{dt} &=& [I][G_0] - [O_1][G_1] - [O_1][O_2] + [O_1][X]\\
\frac{d\left[O_2\right]}{dt} &=& [O_1][G_1] - [O_1][O_2] + [O_2][X]\\
\frac{d [X]}{dt} &=& 2[O_1][O_2] - [O_1][X] - [O_2][X],
\end{eqnarray*}

Following the same arguments in Section 2.1 and Section 2.2, we can eliminate $[G_0]$ and $[X]$ using the conservation laws and we have
\begin{eqnarray}
\frac{d [I]}{dt}&=&-[I] ([I]+G_0-I), \\
\frac{d[G_1]}{dt}&=&-[O_1][G_1]\\
\frac{d\left[O_1\right]}{dt}&=&\left[I\right]\left(\left[I\right]+G_0-I\right)-\left[O_1\right]\left[G_1\right]-\left[O_1\right]\left[O_2\right]\nonumber \\
&& +\left[O_1\right]\left(I-\left[I\right]-\left[O_1\right]-\left[O_2\right]\right),\\
\frac{d\left[O_2\right]}{dt}&=&\left[O_1\right][G_1]-[O_1][O_2]+\ \left[O_2\right](I-[I]-[O_1]-[O_2])
\end{eqnarray}

At first, the upconversion dominates the overall dynamics because the initial points are zero on $[O_1]-[O_2]$ plane, where the vector field of consensus network vanishes. 
After the dynamics of upconversion network reaches its stable points, the entire dynamics switches to the consensus network. 
As was described in Section 2.2, if $I$ is small and $[I]$ goes to 0, the system reduces to the following consensus network:
\begin{eqnarray}
\frac{d\left[O_1\right]}{dt} &=& \left[O_1\right]\left(I-\left[O_1\right]-2\left[O_2\right]\right),\\
\frac{d\left[O_2\right]}{dt} &=& \left[O_2\right](I-2\left[O_1\right]-\left[O_2\right]).
\end{eqnarray}

On the other hand, if $I$ is large enough and $[I]$ goes to $I-G_0$, the system becomes
\begin{eqnarray}
\frac{d\left[O_1\right]}{dt} &=& \left[O_1\right]\left(G_0-\left[O_1\right]-2\left[O_2\right]\right),\\
\frac{d\left[O_2\right]}{dt} &=& \left[O_2\right](G_0-2\left[O_1\right]-\left[O_2\right]).
\end{eqnarray}

As is shown in Fig.\ref{CN-UN}(right), the final state depends on state of the system when the switching from upconversion and consensus network occurs, which explains how the combination of consensus network and upconversion network work as a translator of concentration of input chemical species $I$.

\section{In-silico demonstration of two type of translators} \label{in-silico}
In this section, we demonstrate that how the chemical reaction networks described in the previous can be implemented by DSD reaction and behave under realistic experimental setup.
Specifically, we consider two types of chemical reaction network, translator A and tranlator B.
Translator A is a composite of consensus network and upconverter described in Section \ref{geometry} and translator B is composed of upconverters and downconverters.

\subsection{Translator A with upconverters and expanded consensus network}\label{translator A}
\begin{figure}[htbp]
\begin{center}
\includegraphics[width=1.0\hsize]{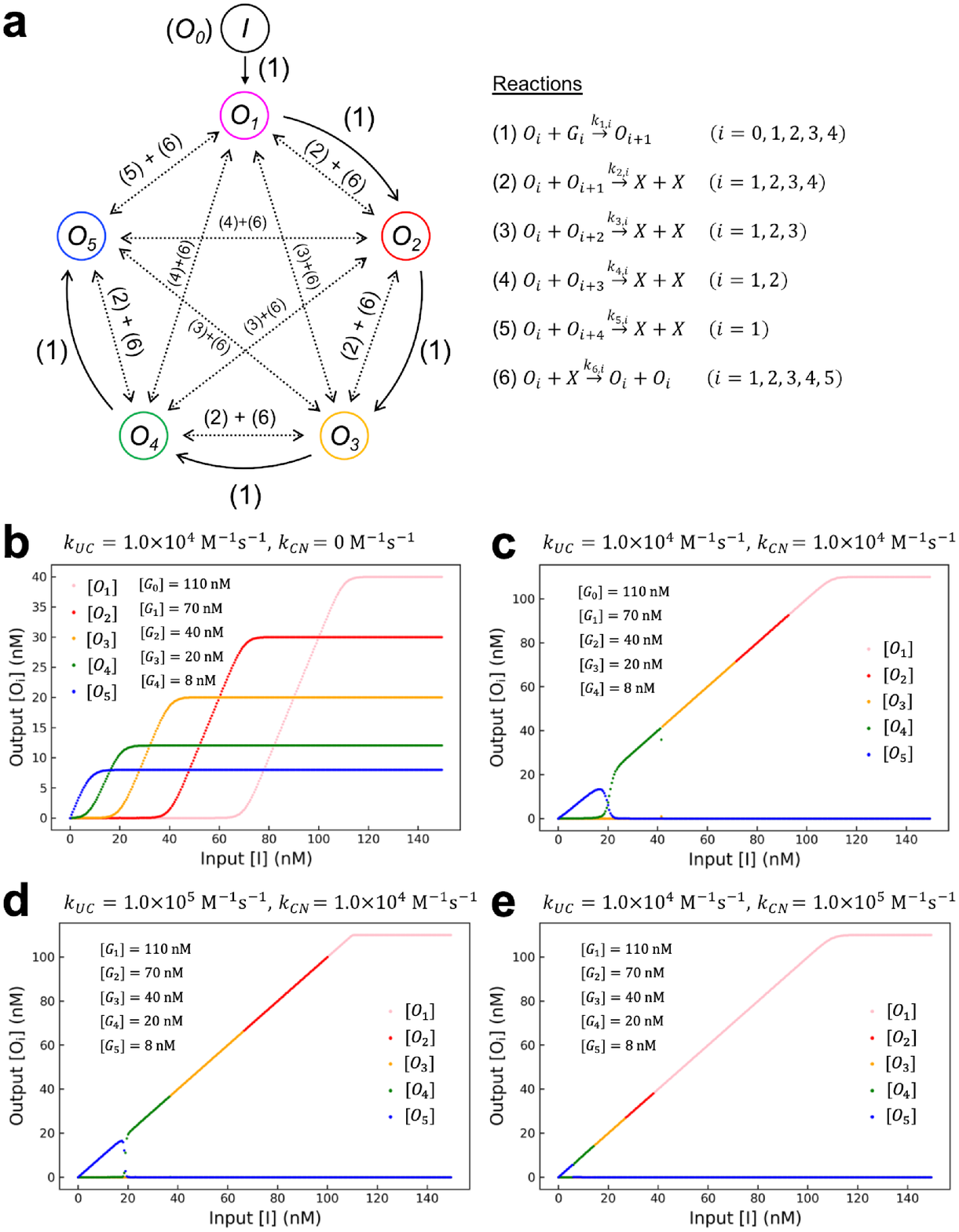}
\caption{(a)Architecture of translator A. Input strand species is represented by $O_0$. 
(b)Remaining strand composition after 20 hours operations of translator A under the condition of $k_{UC} = 1.0 \times\ 10^4~ {\rm M}^{-1}{\rm s}^{-1}, k_{CN} = 0~ {\rm M}^{-1}{\rm s}^{-1}$, 
(c)$k_{UC} = k_{CN} = 1.0 \times\ 10^4 ~ {\rm M}^{-1}{\rm s}^{-1}$, 
(d)$k_{UC} = 1.0\ \times\ 10^5 ~ {\rm M}^{-1}{\rm s}^{-1}$ and $k_{CN} = 1.0\ \times\ 10^4 ~ {\rm M}^{-1}{\rm s}^{-1}$, 
(e)$k_{UC} = 1.0\ \times\ 10^4 ~ {\rm M}^{-1}{\rm s}^{-1}$ and $k_{CN} = 1.0\ \times\ 10^5 ~ {\rm M}^{-1}{\rm s}^{-1}.$
}
\label{Translator A}
\end{center}
\end{figure}

The architecture of translator A and corresponding master equations are shown in Fig.\ref{Translator A}a. 
Here we consider the case of 5 outputs species as an example, although the number of outputs can be arbitrarily increased as explained later in this paper. 
Reaction (1) is upconverters, and reaction (2) - (6) compose an extended consensus network. 
$O_i, G_i$, and $X$ represent output strands, gate strands, and buffer strand respectively.
While the original consensus network by Chen \cite{Chen} involves two species, the presented consensus network in this paper involves all of the output strands (here we exemplify the case of 5 output species) sharing the single buffer strand $X$. 
$k_{j,i}$ is a reaction rate constant, where $j$ indicates the reaction equation numbers, and $i$ is the indexing of related strand species ($k_{i,j}$ is defined only for $i$ listed in the parenthesis following each equation). 
The reaction dynamics follows a set of differential equations shown below.
\begin{eqnarray}
\frac{d[G_i]}{dt} &=& -k_{1,i}[O_i][G_i],\quad (i=0, 1, 2, 3, 4)\\
\frac{d[O_i]}{dt} &=& k_{1,i-1}[O_{i-1}][G_{i-1}] - k_{1,i}[O_i][G_i] - k_{2,i}[O_i][O_{i+1}] - k_{3,i}[O_i][O_{i+2}]\\
&& - k_{4,i}[O_i][O_{i+3}] - k_{5,i}[O_i][O_{i+4}] + k_{6,i}[O_i][X],\quad (i=0,1,2,3,4,5) \nonumber \\
\frac{d[X]}{dt} &=& \sum_{i=1}^4 2 k_{2,i}[O_i][O_{i+1}]
                  + \sum_{i=1}^3 2 k_{3,i}[O_i][O_{i+2}]
                  + \sum_{i=1}^2 2 k_{4,i}[O_i][O_{i+3}]\\
                &&+ 2k_{5,1}[O_1][O_5]  
                  - \sum_{i=1}^5 k_{6,i}[O_i][X]  \nonumber
\end{eqnarray}

Rate constant $k$ for undefined $i$ is regarded as zero. 
For simplicity, we assume that all $k_{1,i}$ are the same and also $k_{2,i}, k_{3,i}, k_{4,i}$ and $k_{5,i}$ are the same. 
Here we redefine the rate constants of the upconverter ($k_{1,i}$) and the rate constants inside the consensus network ($k_{2,i}, k_{3,i}, k_{4,i}, k_{5,i}$) as $k_{UC}$ and $k_{CN}$ respectively. 
First, we assume all bimolecular rate constants to be $1.0\ \times\ 10^4 {\rm M}^{-1}{\rm s}^{-1}$ unless otherwise indicated.
This value is in a realistic range of rate constant for bimolecular DSD reaction \cite{Zhang}. 
This point is explained in more detail later. 
Note that we do not incorporate reverse reaction for each formal reaction, because DSD reaction can be designed to suppress reverse reaction.

We have already analyzed the mechanism of translator A in Section \ref{geometry}, we can also understand its functionality from viewpoint of chemical reaction.
We consider functions of the upconverters and the consensus network separately. 
First, in the case that the consensus network does not takes place (reaction (2) - (6) are absent), upconverters (reaction (1)) convert input strands into larger indexed output strands while consuming gate strands ($G_i$) until input strands or gate strands are used up. 
Fig.\ref{Translator A}b shows the output strand concentrations versus initial input strand concentration provided only by the upconverters after 20 hours reaction time. 
The concentrations of gate strands are indicated on the plot. 
As shown in Fig.\ref{Translator A}b, the major strand species changes successively corresponding to the input concentration. 
This behavior is derived from the gradient of the gate strand concentration. 
As input strands increase from zero, $G_4$ is used up at a certain input level so that the subsequent increase of input strand causes accumulation of $O_4$. 
In this manner, the major strand species switches in turn. 
This switching behavior is essential to single out the major strand species by consensus network as explained below.

In the next step, we consider the functions provided by both the upconverters and the consensus network. 
Along with the successive production of output strands driven by upconverters, the consensus network (reaction (2)-(6)) leaves the major strand species. 
The consensus network in our translator is extended from the original consensus network of Chen \cite{Chen} such that more than two species can make consensus. 
Our consensus network is composed of ten non-catalytic reactions (reaction (2)-(5) for each $i$) and five catalytic reactions (reaction (6) for $i=1, 2,\ldots, 5$). 
$X$ is a buffer signal strand which is shared by all the consensus network reactions. In the consensus network reactions, all the output strands react each other first to generate buffer strand $X$ by non-catalytic reactions (2) - (5). 
Subsequently, buffer strands are consumed by catalytic reactions (6). 
The reaction rate of the catalytic reactions is in proportion to the concentration of each output strand species, and as a result the major output strand population grows faster and finally dominates. 
The remaining strand composition after 20 hours is shown in Fig.\ref{Translator A}c. 
In a wide range of input strand concentration, only single output strand becomes dominant. 
The output strand concentration increases proportionately as input concentration increases, and subsequently output strand species switch at certain input concentrations where each gate strand is used up. 
Finally output strand concentration saturates when the gate strand $G_0$ is used up. 
This switching behavior is exactly what realizes the function of our translator.

Now we try to qualitatively understand this switching behavior. 
The behavior is governed by the concentrations of the gate strands and rate constants of reactions. 
First, as already mentioned, the gate strand concentrations should have gradient in order to switch the major strand corresponding to the input strand concentration, and the switching values, that indicates the input concentrations on which the output strand switch from one strand species to another strand species, are mainly determined by the concentration of each gate. 
Regarding the rate constants, we focus on the relative ratio of rate constants because absolute values only change the timescale in which the translator works. 
Relatively higher rate constants of the upconverters ($k_{UC}$) than those of consensus network ($k_{CN}$) result in more drastic switching behavior as shown in Fig.\ref{Translator A}d, because the output strand concentrations more directly follow the concentrations prepared by the upconverters as shown in Fig.\ref{Translator A}b. 
On the other hand, with a higher-rate consensus network, more strands are converted into larger-indexed strand species than with a lower-rate consensus network. 
This is because an amount of the larger-indexed strand species always exceeds than that of the lower-indexed one due to the gradient of the gate strand concentration. 
As a result, the switching values shift to lower input concentration with higher $k_{CN}$. 
Therefore, the dynamic range of the translator can be adjusted by both the gate strand concentrations and rate constants of each reaction. 
It should be noted that isolation of a single output strand is not so clear in the lower input range while it's clear in the higher input range. 
This is because the reaction rate is slower with the lower input due to lower reactants' concentration so that the time required to reach a steady state is longer than that with higher input concentration.
The mathematical analysis of transient dynamics of the translator is described in detail in appendix section.

\begin{figure}[htbp]
\begin{center}
\includegraphics[width=1.0\hsize]{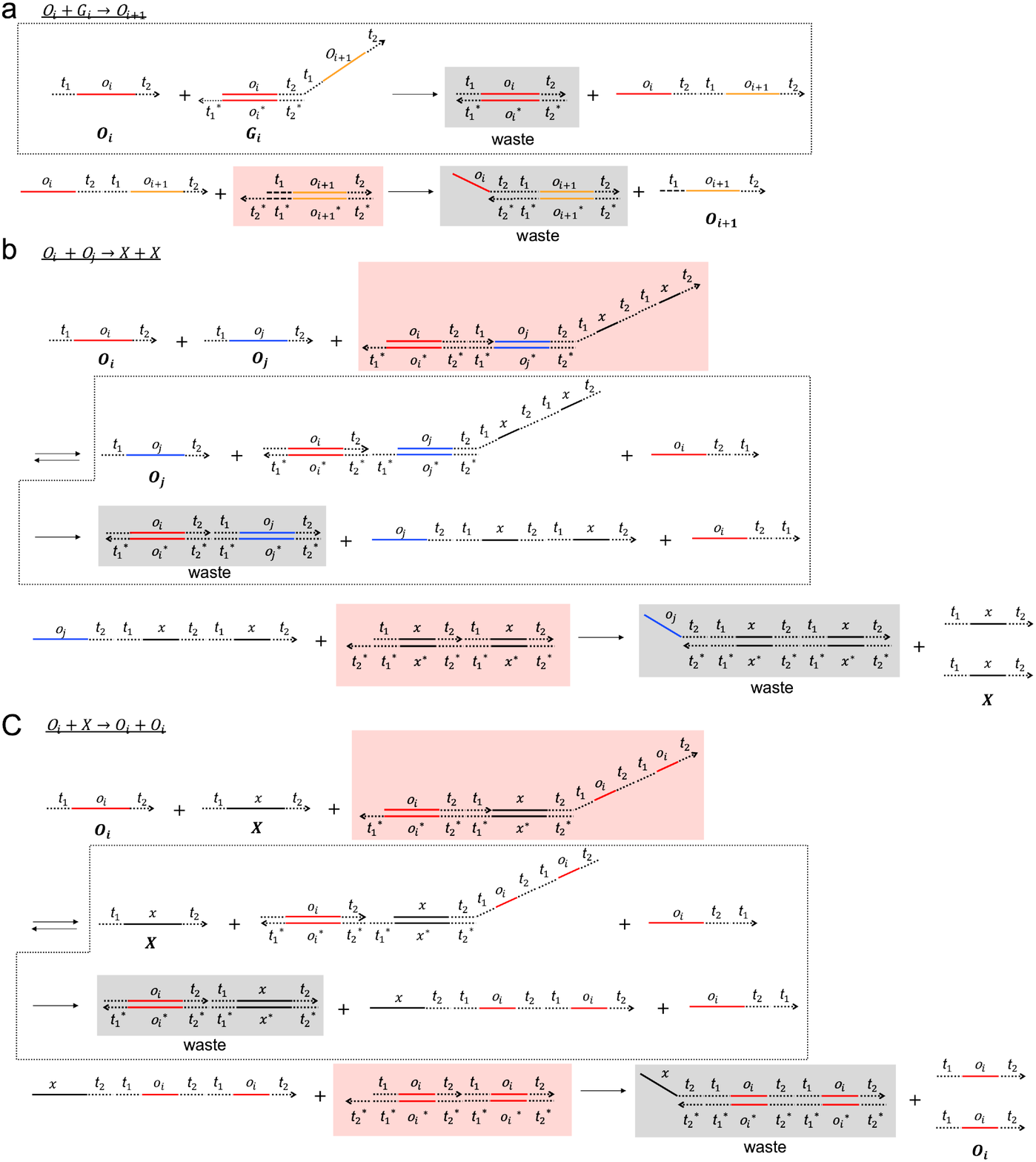}
\caption{Biochemical implementation of translator A. Strands highlighted in pink are auxiliary strands for realizing desired functions. Strands highlighted in gray are waste strands, which no longer react with other strands.}
\label{Biochemical Translator A}
\end{center}
\end{figure}

Biochemical implementation of translator A is shown in Fig.\ref{Biochemical Translator A}, which is based on the previous work by Soloveichik et al \cite{Soloveichik}. 
DNA sequences are represented by arrows which direct from 5' to 3'. 
Each of DNA strands included in the reaction equations comprises two types of sequence domain: a representative domain of each strand species represented by a lowercase letter, and toehold domains represented by t, by which a DSD reaction can be initiated. 
In addition to the strand species indicated in the reaction equations in Fig.\ref{Translator A}a, there are other strand species involved in the reactions, called auxiliary strands, which are highlighted by the pink boxes in Fig.\ref{Biochemical Translator A}. 
We assume that there is an excessive amount of the auxiliary strands. 
Thereby we can approximate all formal reactions shown in Fig.\ref{Translator A}a to be bimolecular reactions, because only bimolecular elementary reactions indicated by the dotted square lines in Fig.\ref{Biochemical Translator A} are rate-limiting steps with non-excess amounts of reactants. 
The gray boxes in Fig. \ref{Biochemical Translator A} indicate waste strands which do not participate in any subsequent reactions including the reverse reaction of each elementary reaction. 
Although the reverse reactions occur slightly, the reaction rates of the reverse reactions are so slow to be negligible.

The kinetics of DSD reaction can be well-predicted by mathematical model, as shown by the work of Zhang and Winfree, in which the mathematical model showed good agreement with experimental results within an order of magnitude \cite{Zhang}. 
According to their work, a rate constant of a DSD reaction can be controlled by the number of bases and GC contents of the toeholds over 6 orders of magnitude ($1.0 - 1.0\ \times\ 10^6 {\rm M}^{-1}{\rm s}^{-1}$), under an assumption that there is no secondary structure in the toehold domain. 
Therefore, $10^4$ and $10^5 {\rm M}^{-1}{\rm s}^{-1}$ we used in the computational analysis is a plausible value for a rate constant of a DSD reaction. 

\subsection{Translator B with upconverters and downconverters}
\begin{figure}[H]
\begin{center}
\includegraphics[width=1\hsize]{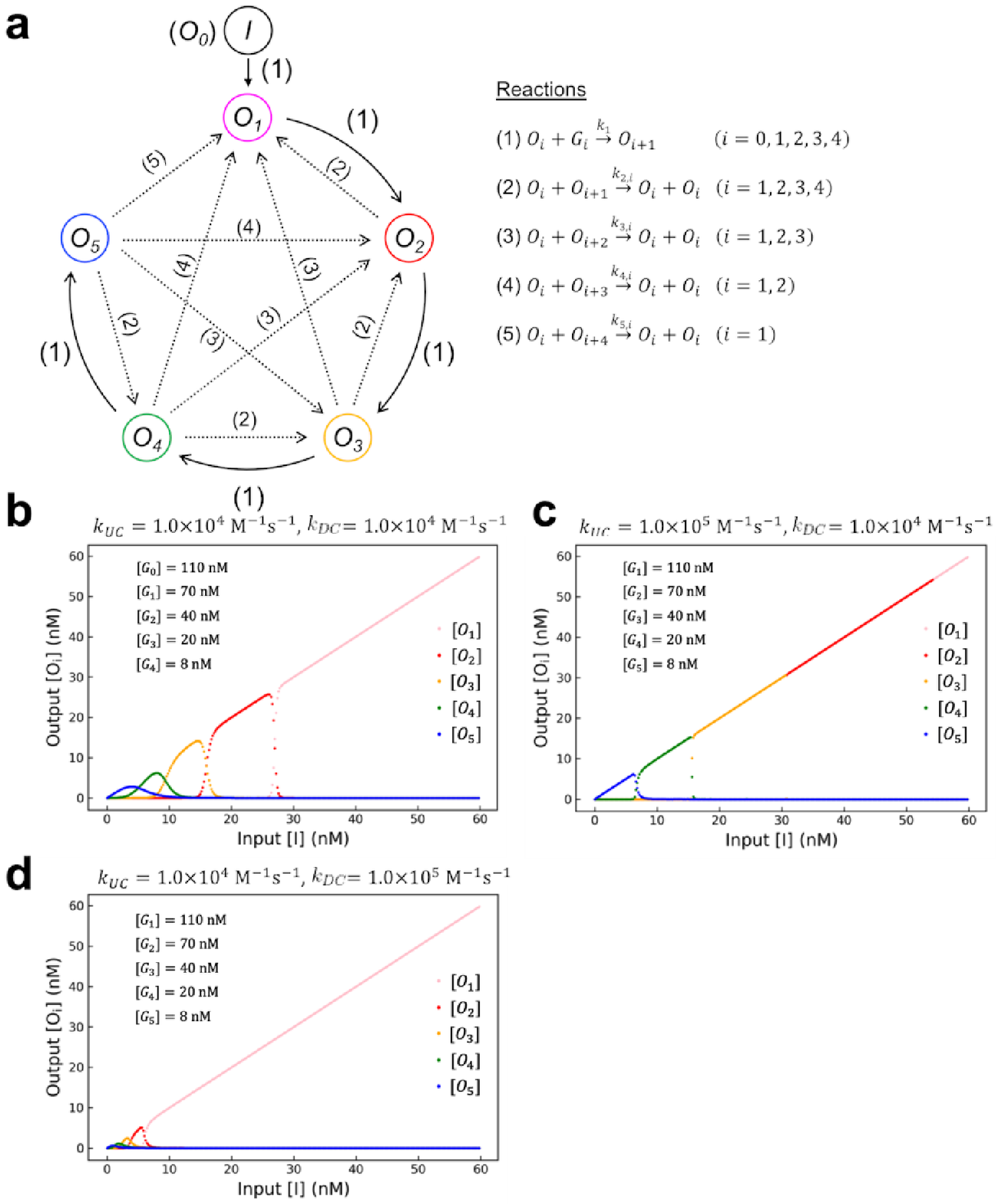}
\caption{(a)Architecture of translator B. Input strand species is represented by $O_0$. 
(b)Remaining strand composition after 20 hours operations of translator B under the condition of $k_{UC} = k_{CN} = 1.0 \times\ 10^4~ {\rm M}^{-1}{\rm s}^{-1}$, 
(c)$k_{UC} = 1.0 \times\ 10^5~ {\rm M}^{-1}{\rm s}^{-1}$ and $k_{CN} = 1.0 \times\ 10^4 ~ {\rm M}^{-1}{\rm s}^{-1}$, 
(d)$k_{UC} = 1.0\ \times\ 10^4 ~ {\rm M}^{-1}{\rm s}^{-1}$ and $k_{CN} = 1.0\ \times\ 10^5 ~ {\rm M}^{-1}{\rm s}^{-1}$.
}
\label{Translator B}
\end{center}
\end{figure}

\begin{figure}[H]
\begin{center}
\includegraphics[width=1.0\hsize]{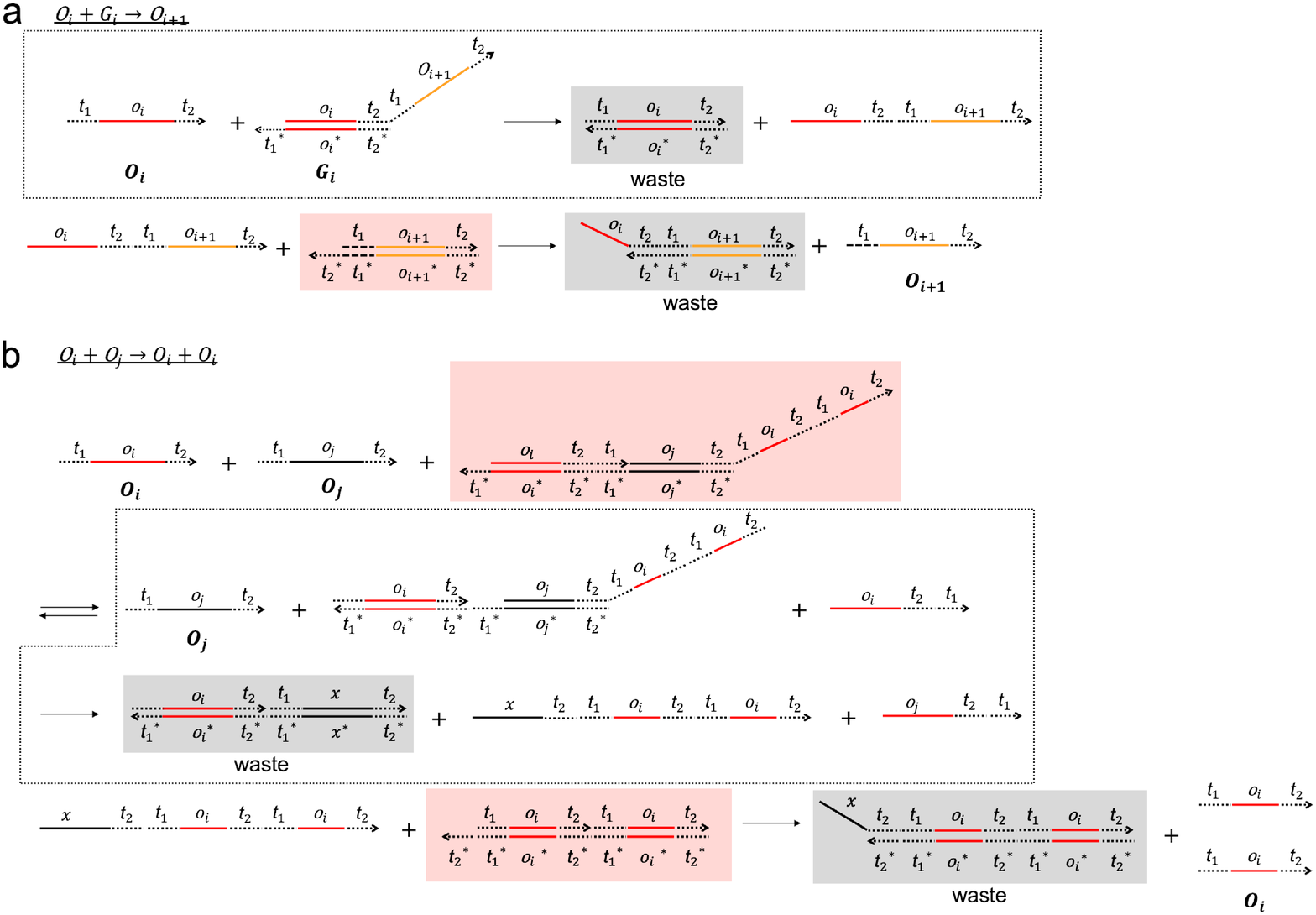}
\caption{Biochemical implementation of translator B. 
Strands highlighted in pink are auxiliary strands for realizing desired functions. 
Strands highlighted in gray are waste strands, which no longer react with other strands.}
\label{Biochemical Translator B}
\end{center}
\end{figure}

We also propose translator B which does not have consensus network but still have a similar network structure as shown in Fig.\ref{Translator B}a. 
Biochemical implementation of translator B is shown in Fig. \ref{Biochemical Translator B}.
Translator B is composed of upconverters (reaction (1)), the same as those of translator A, and also the downconverters (reaction (2) - (5)) which are unique to translator B. 
The downconverters convert larger-indexed output strand species into smaller-indexed output strand species.  
There are two major differences between the consensus network and the downconverters. 
First, the downconverters do not involve any buffer strands, so output strands directly react each other. 
Second, the downconverters compete with the upconverters, whereas the consensus network involves competitions among the members of the consensus network for winning the majority. 
Therefore, in translator B, the ratio of reaction rates of upconverters and downconverters have an essential role for determining the switching behavior.
The reaction dynamics of translator B follow a set of differential equations as shown below.
\begin{eqnarray}
\frac{d[G_i]}{dt} &=&-k_{1,i}[O_i][G_i],\quad (i=0, 1, 2, 3, 4)\\
\frac{d[O_i]}{dt} &=& k_{1,i-1}[O_{i-1}][G_{i-1}]
                     + k_{2,i-1}[O_{i-1}][O_i]
                     + k_{3,i-2}[O_{i-2}][O_i]\\ 
                  && + k_{4,i-3}[O_{i-3}][O_i]
                     + k_{5,i-4}[O_{i-4}][O_i]
           		     - k_{1,i}[O_i][G_i]
           		     - k_{2,i}[O_i][O_{i+1}] \nonumber \\
           		  && - k_{3,i}[O_i][O_{i+2}] 
           		     - k_{4,i}[O_i][O_{i+3}]
           		     - k_{5,i}[O_i][O_{i+4}], \quad (i=0, 1, 2, 3, 4, 5) \nonumber
\end{eqnarray}
Here, we assume that rate constants of upconverters ($k_{1,i}$) are identical and that the rate constants of downconverters ($k_{2,i}, k_{3,i} , k_{4,i} , k_{5,i}$) are identical, so $k_{1,i}$ is represented by $k_{UC}$ and $k_{2,i}, k_{3,i}, k_{4,i}, k_{5,i}$ are represented by $k_{DC}$.
Fig.\ref{Translator B}b shows the remaining strand composition after 20 hours of translator B operation with $k_{UC}: 1.0\ \times\ 10^4 ~ {\rm M}^{-1}{\rm s}^{-1}$ and $k_{DC}: 1.0\ \times\ 10^4 ~ {\rm M}^{-1}{\rm s}^{-1}$. 
Even with the same gate strand composition, the switching values are different from that of translator A. 
The switching values of translator A are determined mainly by the gate strand composition and partly affected by the rate constants. 
However, the switching values of translator B are strongly dependent on the rate constants. Fig.\ref{Translator B}c and Fig.\ref{Translator B}d shows the remaining strand compositions with different rate constants. 
With higher $k_{UC}$, the switching values shift to larger input concentrations, while with higher $k_{DC}$ the switching timings shift to lower input concentrations. 
This behavior can be simply interpreted as a result from the competition of upconverters and downconverters. 

\subsection{Comparison of translator A and B}
There are two points to be considered when we compare translator A and B: switching behavior and biochemical implementation cost. 
First, the switching behavior of translator B is more sensitive to the rate constants than that of translator A. 
As already mentioned, this sensitivity is a result of the competition between the upconverters and the downconverters. 
This feature provides a tunability of a dynamic range of the input strand concentration translator. 
On the other hand, the sensitivity can also be interpreted as instability of the behavior of the translator. 
Therefore, both concentration translators should be employed properly according to requirements from application stand points. 
It should be noted that the switching behavior is also controlled by the gate strand concentrations. 
If the maximum concentration of DNA strands in a reaction system (in other words, biochemical resources) is constant, an increase of each gate strand concentration limits the number of output strand species to be processed while retaining substantial concentration. 
Therefore, in the present study, we set the gate strand concentration at constant when calculating the translator dynamics.
Next, we discuss the biochemical implementation cost of both translators. Here, the term "biochemical implementation cost" simply means the the number of DNA species involved in the chemical reaction networks. 
Table \ref{comparison} shows a comparison of the number of DNA strand species required to implement each translator circuit when the number of the output strand is $N$. 
The total biochemical cost (DNA concentration) is predominantly determined by the concentrations of the auxiliary strands, because they should be larger than other strand species to keep the reaction system the set of bimolecular reactions as described by the reaction equations in Fig. \ref{Translator A}a and Fig. \ref{Translator B}a. 
Note that the absolute number of the auxiliary strand species depends on the specific biochemical implementation, while we can still relatively compare the number of strand species in both translators. 
Translator B requires a smaller number of auxiliary strand species to be biochemically implemented, because it does not involve buffer strands X, which are required by translator A. 
However, if $N$ is large enough, the term of $N^2$ becomes more dominant. Thus, both translators are comparable in terms of the cost for biochemical implementation. 


\begin{table}[htbp]
\begin{center}
\caption{Number of DNA strand species to biochemically implement each translator circuit.}
  \begin{tabular}{ccc} 
    \hline
    {\bf Strand species} & {\bf Translator A} & {\bf Translator B} \\ 
    \hline
    Output strand        &     $N$            & $N$   \\ 
    Buffer strand        &       1            &   0   \\
    Gate strand          &     $N$            & $N$   \\
    Auxiliary strand     &     $N^2 + 2N$     & $N^2$ \\ 
    \hline
                         &                    & $N$: number of output strand \\
  \end{tabular}
  \label{comparison}
\end{center}
\end{table}

\section{Conclusion}
In the present work, we proposed muti-species consensus networks by chemical reaction networks and showed that they can perform as concentration-to-strand translators.
The dynamics of the translator was understood as heteroclinic network from the viewpoint of nonlinear dynamical systems.  
It was successfully demonstrated that two types of translators output a unique output strand species corresponding to a value of the input strand concentration. 
Translator A and B showed a slightly different behaviors which offers tunable options depending on applications. 
Our translators map analog concentration signal to digital information, that is, set of multiple DNA strands.
This functionality provides easy-to-use biomarkers which are potentially useful for on-site personal healthcare systems since no costly fluolescence-based techniques are required.
Such direction of research is left to futre work.

\bibliography{references}

\end{document}